\documentclass[twocolumn]{jpsj2}

\def\QGS{Q_{\rm GS}}
\def\QHAM{Q_{\rm Ham}}
\def\Ms{M_{\rm s}}
\def\Mtot{M_{\rm tot}}
\def\Mpa{M_{\rm p1}}
\def\Mpb{M_{\rm p2}}
\def\Mpc{M_{\rm p}}

\def\JA{J_{\rm A}}
\def\JF{J_{\rm F}}

\def\v#1{\mib #1}

\def\runtitle{Magnetization Plateaux in  Random Frustrated S=1/2 Heisenberg Chains
}

\def\runauthor{Kazuo {\sc Hida}}

\title
{
Magnetization Plateaux in  Random Frustrated S=1/2 Heisenberg Chains
}

\author
{Kazuo {\sc Hida}
\footnote{e-mail: hida@phy.saitama-u.ac.jp}}

\inst
{
Department of Physics, Faculty of Science,\\ Saitama University, Saitama 338-8570
}

\recdate
{
\today
}

\abst
{
The $S=1/2$ frustrated Heisenberg chains  with bond alternation is known to exhibit a magnetization plateau at half of the saturation magnetization $\Ms$ accompanied by the spontanuous translational symmetry breakdown. The effect of randomness on the magnetization process of this model is investigated. First we consider the mixture of the two kinds of chains both of which possess the $\Ms/2$-plateau in the common interval of the magnetization field. The plateau at $\Ms/2$ is found to vanish immediately if the randomness is switched on in agreement with Totsuka's prediction. The small plateau also appears near the saturation field due to the localization of inverted spins around the minority bond. On the other hand, if the stronger bond is replaced by the ferromagnetic bonds randomly, the randomness induced fractional plateau appears as in the nonfrustrated case. The  plateau at $\Ms/2$ does not simply vanish but shifts and splits into two smaller plateaux. The magnetization on this plateau varies nonlinearly with $1-p$. The physical origin of this behavior is explained based on the strong coupling picture.
}
\kword
{
random Heisenberg chain, bond alternation, frustration, DMRG, magnetization plateau
}

\begin{document}
\maketitle


\section{Introduction}
In the recent studies of quantum many body problem, magnetization plateaux in one dimensional Heisenberg chains are attracting much attention as the field induced spin gap states. On the plateau, some spins are partly quenched by the magnetic field while the remaining spins form a spin gap state\cite{hk,tone,tone1,totsuka1}.  

Oshikawa, Yamanaka and Affleck\cite{oya} proposed the  necessary condition for the magnetization plateaus as 
\begin{equation}
\QGS(S-M)=Z\equiv \mbox{integer}
\label{oyac}
\end{equation}
where $\QGS$ is the spatial periodicity of the magnetic ground state, $S$ is the magnitude of the spin, and $M$ is the magnetization per site. It should be noted that $\QGS$ is not always equal to the spatial periodicity of the Hamiltonian $\QHAM$. If the translational symmetry is spontanuously broken,  $\QGS=z\QHAM$ where $z$ is an integer.

Recently,  the effect of randomness on the magnetization plateaux of one-dimensional quantum spin systems has been studied theoretically\cite{cabra,khr,totsukar}. Among them, the present author found the randomness induced plateau in the mixture of  $S=1/2$ antiferromagnetic-antiferromagnetic alternating Heisenberg chain and  $S=1/2$ ferromagnetic-antiferromagnetic alternating  Heisenberg chain\cite{khr} which models the compound  (CH$_3)_2$CHNH$_3$Cu(Cl$_x$Br$_{1-x})_3$ studied by Manaka and coworkers\cite{manaka}. Similar observation has been made by Cabra and coworkers\cite{cabra} for random $q$-merized chains. 

On the other hand, Totsuka\cite{totsukar} discussed the effect of randomness on the magnetization plateaux which are already present in periodic systems. He concluded that the plateau induced by the {\it imposed} periodicity ($\QGS=\QHAM$) is stable against weak randomness whereas the plateau induced by the spontanuous translational symmetry breakdown (STSB) ($\QGS \ne \QHAM$) is unstable against randomness due to Imry-Ma effect\cite{imryma}. 

Therefore we know that randomness induces the magnetization plateau on one hand  and on the other hand it destroy the plateau. It is the purpose of the present work to investigate the interplay of these two apparently contradicting aspects of randomness effect on the magnetization plateau phenomenon using the density matrix renormalization group (DMRG) method\cite{wh1}.

This paper is organized as follows. In the next section, the model Hamiltonian is defined. Frustrated random chains  with bond alternation are investigated in section 3.1  and frustrated chains  with bond alternation and random sign strong bonds are studied in section 3.2 using the DMRG method.   The physical interpretation of the numerical results are also given. The final section is devoted to a summary and discussion.

\section{Model Hamiltonian}

\begin{figure}
\centerline{\includegraphics[width=80mm]{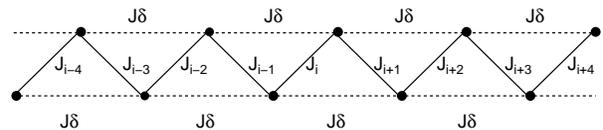}}
\caption{Heisenberg chain with next-nearest-neighbour interaction. The nearest neighbour interaction is spatially modulated.}
\label{zchain}
\end{figure}

As a simplest model which exhibits a plateau accompanied by STSB, we consider  the $S=1/2$ Heisenberg chain with next-nearest-neighbour interaction depicted in Fig. \ref{zchain}. The Hamiltonian is given by,

\begin{figure}
\centerline{\includegraphics[width=60mm]{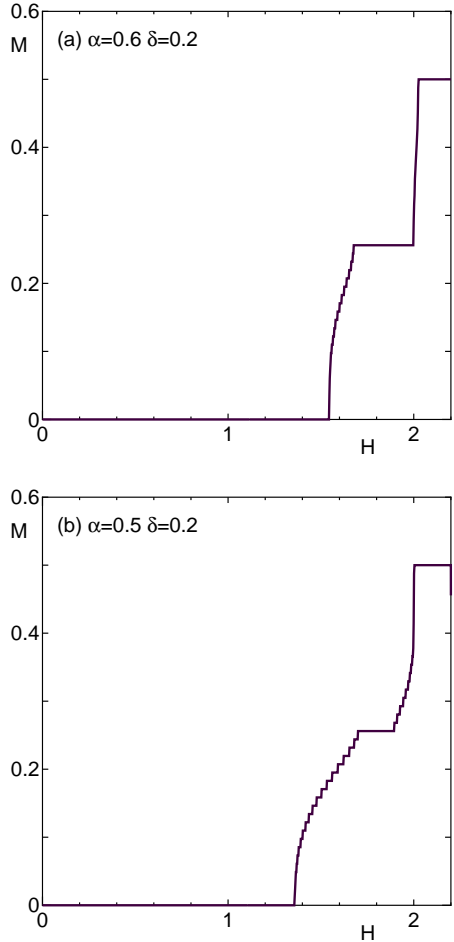}}
\caption{The magnetization curves of the pure frustrated random Heisenberg chains with  $\delta=0.2$, (a) $\alpha=0.5$ and (b) $\alpha=0.6$. The system size is $N=82$.}
\label{figpure}
\end{figure}

\begin{equation}
\label{ham1}
H = \sum_{i=1}^{N-1}J_i \v{S}_{i} \v{S}_{i+1}+ \sum_{i=1}^{N-2}J\delta \v{S}_{i} \v{S}_{i+2}
\end{equation}
where the bond alternation and randomness is introduced in the distribution of $J_i$. In the present work, we consider the following two types of randomness: 
\begin{enumerate}
\item  Random bond alternation

\begin{equation}
J_i=
\left\{
\begin{array}{ll}
J(1+(-1)^i\alpha_1) & \mbox{: probability}\ p \\
J(1+(-1)^i\alpha_2) &  \mbox{: probability}\ 1-p
    \end{array}
\right.
\end{equation}

\item  Random sign strong bonds
\begin{eqnarray}
&&J_{2i-1} =J > 0 \nonumber \\
&&J_{2i}=
\left\{
\begin{array}{ll}
\JA (> J) & \mbox{: probability}\ p \\
\JF (<0) &  \mbox{: probability}\ 1-p
    \end{array}
\right.
\end{eqnarray}
\end{enumerate}
In the following, we set $J=1$ to define the energy unit.

In the pure case ($\alpha_1=\alpha_2 \equiv \alpha$), this model has a plateau at the magnetization $M=\Ms/2$   for appropriate values of $\alpha$ and $\delta$\cite{tone,tone1,totsuka1} where $\Ms$ is the saturation magnetization per site ($\Ms \equiv 1/2$ in the present model). To explain this plateau, we have to set $\QGS=2\QHAM$ in Eq. (\ref{oyac}). Therefore this plateau is accompanied by STSB. Figures \ref{figpure}(a) and (b) show typical examples of such cases with $\delta=0.2$ and (a) $\alpha=0.5$ and (b) $\alpha=0.6$ calculated by the DMRG method.  

\section{Numerical Results}
\subsection{Case 1: Frustrated Heisenberg chain with random bond alternation }

\begin{figure}
\centerline{\includegraphics[width=60mm]{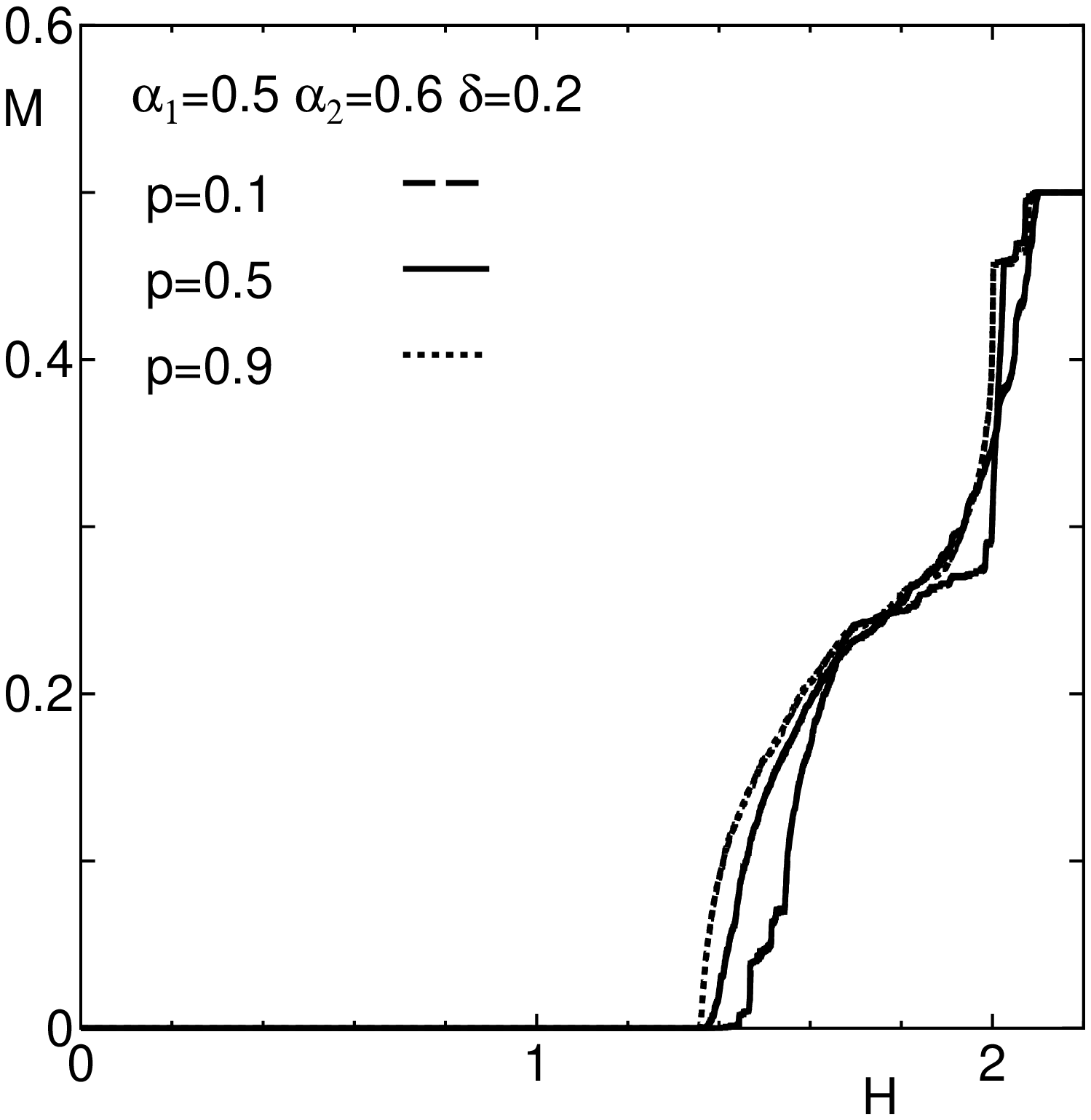}}
\caption{The magnetization curves of the random frustrated Heisenberg chains with  $\delta=0.2$, $\alpha_1=0.5$ and $\alpha_2=0.6$ for various values of $p$. The system size is $N=82$ and average is taken over 40 samples.}
\label{figran}
\centerline{\includegraphics[width=60mm]{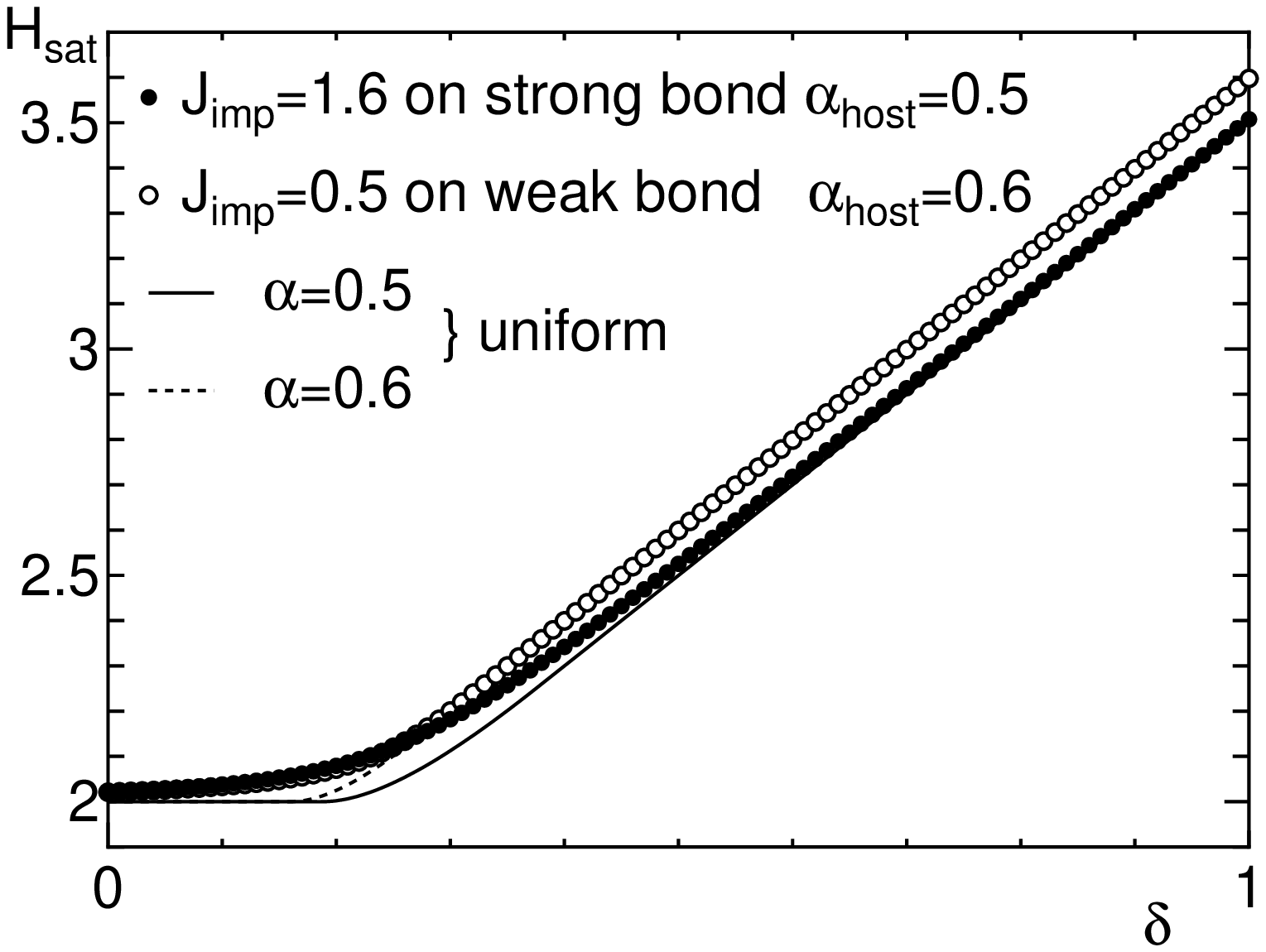}}
\caption{The saturation fields $H_{\rm sat}$ of the frustrated Heisenberg chains with an impurity bond with $N=82$. The filled (open) circles represent the saturation field of the chain with $\alpha=0.5 (0.6) $ with a single strong (weak) bond replaced by the bond with $J_{\rm imp} = 1.6$ (0.5).  The solid and dotted lines are the saturation fields of the pure chains with $\alpha =0.5$ and 0.6, respectively. }
\label{fsat}
\centerline{\includegraphics[width=60mm]{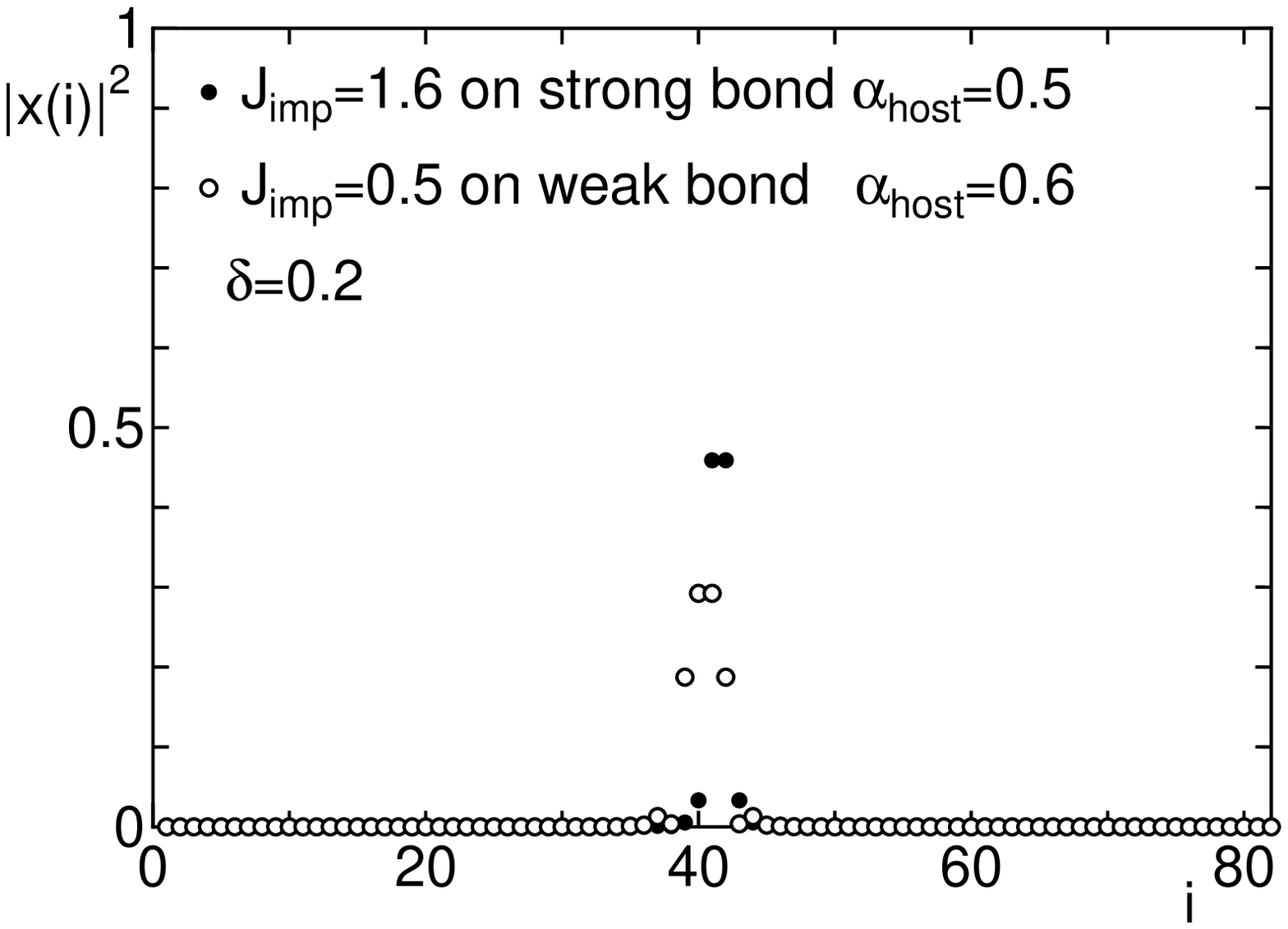}}
\caption{The probability density $\mid x(i)\mid^2$  of the inverted spin in the ground state with $\Mtot=N/2-1$ of the frustrated Heisenberg chains with an impurity bond. The chain length is $N=82$. The filled (open) circles represent $\mid x(i)\mid^2$  of the chain with $\alpha=0.5 (0.6) $ with a single strong (weak) bond replaced by the bond with $J_{\rm imp} = 1.6$ (0.5) at $i=41$(40) with $\delta=0.2$.  }
\label{wfsat2s}
\end{figure}
 
From Fig. \ref{figpure}, it is clearly seen that the magnetic field intervals of the plateau region of cases (a) and (b) have finite overlap. In this section, we investigate what happens if these two chains are mixed. The result is shown in Fig. \ref{figran}. Even in the magnetic field interval in which both (a) and (b) have plateau, no plateau appears in the mixed chain. This verifies the fragility of this kind of plateau due to Imry-Ma effect\cite{imryma} as predicted by Totsuka\cite{totsukar}.

It should be also remarked that the plateau with $M=0$ (spin gap) remains stable. This plateau is due to the imposed bond alternation  ($\QGS=\QHAM$) so that this result is also in accord with Totsuka's prediction\cite{totsukar}.

The structure near the saturation is also induced by randomness. It should be remarked that the saturation field  $H_{\rm sat}$ in the mixed chain is larger than those of both pure chains. The saturation field is determined as the energy difference between the fully polarized state ($\Mtot=N/2$) and the state with a single inverted spin ($\Mtot=N/2-1$). Here $\Mtot$ denotes the total magnetization. In the presence of the impurity bond, the inverted spin can be localized around the impurity bond. In Figure \ref{fsat}, the saturation fields of the chains with a single impurity bond with strength $J_{\rm imp}$ in the host chain with bond alternation $\alpha=\alpha_{\rm host}$ is presented for $N=82$ chain. The filled circles represent the saturation field of the chain with $\alpha_{\rm host}=0.5$ and $\delta=0.2$ in which a single strong bond is replaced by the impurity bond with $J_{\rm imp} = 1.6$ corresponding to the strength of the strong bond of the chain with $\alpha=0.6$.  The open circles represent the saturation field of the chain with $\alpha_{\rm host}=0.6$ and $\delta=0.2$ in which a single weak bond is replaced by the impurity bond with $J_{\rm imp} = 0.5$ corresponding to the strength of the weak bond of the chain with $\alpha=0.5$. The saturation field of the pure chain is easily calculated analytically as

\begin{equation}
H_{\rm sat} =
\left\{
\begin{array}{ll}
2 &  \displaystyle{ \delta \leq \frac{1-\alpha^2}{4} }\\
\displaystyle{1+\frac{2\delta}{1-\alpha^2}+\frac{1-\alpha^2}{8\delta}} & \displaystyle{\frac{1-\alpha^2}{4\alpha} \geq \delta \geq \frac{1-\alpha^2}{4}} \\
1+2\delta+\alpha & \displaystyle{\delta \geq \frac{1-\alpha^2}{4\alpha} }\\
\end{array}
\right.
\end{equation}
as plotted in Fig. \ref{fsat} by the  solid and dotted lines for $\alpha =0.5$ and 0.6 with $\delta=0.2$, respectively. It is seen that the saturation field for the chain with an impurity is substantially larger than the pure chains for small $\delta$ where the localized states are formed.  Corresponding probability density of the inverted spin $\mid x(i)\mid^2$  in the ground state with $\Mtot=N/2-1$ is shown in Fig. \ref{wfsat2s} for $\delta=0.2$. This explicitly shows that the inverted spin is localized around the impurity bond. 

 For small but finite density of impurity bonds, the magnetization of the most part of the chain is saturated at the saturation field of the pure chain. However, the portion of the chain proportional to the density of the impurity bond remains unpolarized until the saturation field of the chain with an impurity is reached. Therefore there appears a small plateau at $M=M_sp$ for small $p$ and at $M=M_s(1-p)$ for small $1-p$ as observed in Fig. \ref{figran}. This can be also regarded as a randomness induced plateau.

\subsection{Case 2 : Frustrated chains  with bond alternation and random sign strong bonds}

\begin{figure}
\centerline{\includegraphics[width=60mm]{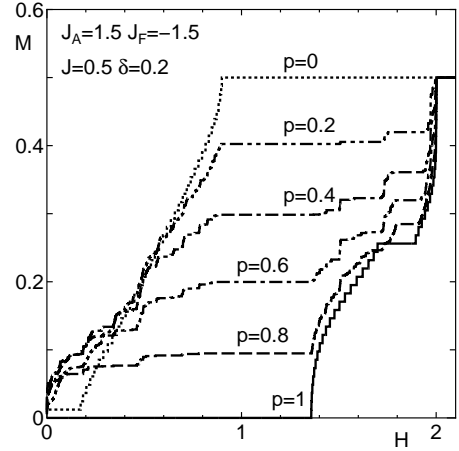}}
\caption{The magnetization curves of the frustrated Heisenberg chains with bond alternation and random sign strong bonds with  $\JA=1.5, \JF=-1.5, J=1.0$ and $\delta=0.2$ for various values of $p$. The system size is $N=82$ and average is taken over 40 samples. The small low field magnetization for $p=1$ is dur to the finite size effect.}
\label{figranf}
\end{figure}
In the absence of frustration $\delta=0$, this model shows the randomness induced plateau at $M=\Ms(1-p)$ as pointed out by the present author\cite{khr}. The same phenomenon takes place in the presence of finite $\delta$. 

\begin{figure}
\centerline{\includegraphics[width=60mm]{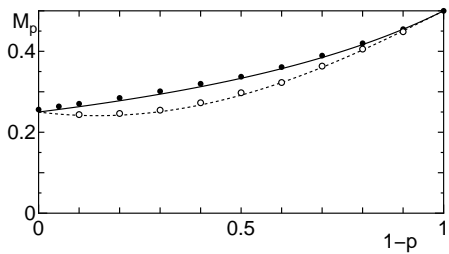}}
\caption{The $1-p$-dependence of the magnetization on the plateau which reduces to the $M=\Ms/2$ plateau in the pure $p=1$ limit for the frustrated Heisenberg chains with bond alternation and random sign strong bonds. The filled and open circles represent $\Mpa$ and $\Mpb$, respectively. The solid and dotted cureves represent the formulae (\ref{mp1}) and (\ref{mp2}), respectively.}
\label{shift}
\end{figure}

The fate of the plateau at $M=\Ms/2$ is different from that of case 1. The plateau no more appears at $M=\Ms/2$ but it does not simply vanish. The plateau splits into two plateaux with magnetizations $\Mpa$ and $\Mpb$  as depicted in Fig. \ref{figranf}. Actually, the lower one ($\Mpb$) of these two plateaux is only visible for relatively large values of $1-p$. Furthermore, the magnetization on both  plateaux depends nonlinearly on $1-p$ as shown in Fig. \ref{shift}.

The nonlinear dependence of  $\Mpa$ and $\Mpb$  on $p$ can be understood based on the strong coupling picture for the plateau state proposed by Totsuka\cite{totsuka1,totsukar} in the following way. On the plateau, all the spin pairs connected by ferromagnetic bonds are polarized. This gives the contribution $\Ms(1-p)$ to the magnetization which is the value for the randomness induced plateau. By these bonds, the chain is cut into segments of finite length. Considering that the neighbouring triplet-triplet or singlet-singlet pairs costs energy\cite{totsuka1,totsukar}, the segments consisting of $2l+1$ strongly coupled antiferromagnetic dimers have stably magnetized state with $l$ triplets pairs. The average number of such segment is $(N/2)(1-p)^2p^{2l+1}$. So that these segments contribute to the magnetization by an amount 

\begin{equation}
\frac{N}{2}\sum_{l=1}^{\infty}l(1-p)^2p^{2l+1}
\end{equation}
On the other hand, for the segments consisting of $2l$ strongly coupled dimers, it is enevitable to have at least one neighbouring triplet-triplet or singlet-singlet pairs. Therefore these segments  have stably magnetized state with $l-1$ or $l$ triplets pairs. The average number of such segment is $(N/2)(1-p)^2p^{2l}$. Therefore they contribute to the magnetization by amounts  
\begin{equation}
\frac{N}{2}\sum_{l=1}^{\infty}(l-1)(1-p)^2p^{2l}, 
\end{equation}
 or 
\begin{equation}
\frac{N}{2}\sum_{l=1}^{\infty}l(1-p)^2p^{2l}
\end{equation}
These summations are easily carried out and the stable plateaux are expected at
\begin{equation}
\Mpa=\Ms(1-p +\frac{p^2}{1+p})
\label{mp1}
\end{equation}
 and  
\begin{equation}
\Mpb=\Ms(1-p +\frac{p^3}{1+p})
\label{mp2}
\end{equation}
These are also plotted in Fig. \ref{shift} by solid and dotted curves. The coincidence between the numerical results and above formulae is fairly good. The slight deviation is attributed to the finite size effect.

\begin{figure}
\centerline{\includegraphics[width=60mm]{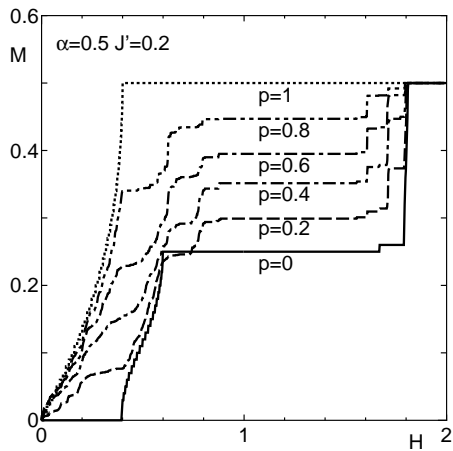}}
\caption{The magnetization curves of the random $q$-merized Heisenberg chains with  $q=4, \alpha=0.5, J'=0.2$ and $J=1.0$ for various values of $p$. The system size is $N=100$ and average is taken over 20 samples.}
\label{magpujo4}
\end{figure}
\begin{figure}
\centerline{\includegraphics[width=60mm]{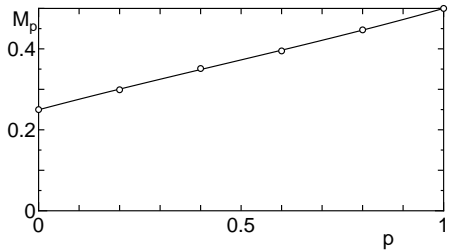}}
\caption{The $p$-dependence of the magnetization on the plateau random $q$-merized Heisenberg chains (open circles) with $q=4$. The solid line is Eq. (\ref{linear}).}
\label{shiftq}
\end{figure}

The nonlinear $p$-dependence of the plateau magnetization is in contrast to the case of the plateau in random $q$-merized chain in which $\delta=0$ and $J_i$ are randomly distributed according to the distribution. 
\begin{equation}
J_i=
\left\{
\begin{array}{ll}
J(1+\alpha_i) & \mbox{: probability}\ 1-p \\
J' &  \mbox{: probability}\ p
    \end{array}
\right.
\label{qmer}
\end{equation}
where $\alpha_i \equiv \alpha,\: (-\alpha)\,$ if $i = q n,\:$ ($ i
\ne q n\,$). The XX model with this exchange distribution is  investigated by Cabra and coworkers\cite{cabra}. Here we present the magnetization curve of the corresponding antiferromagnetic Heisenberg model using the DMRG method in Fig. \ref{magpujo4} for $q=4$. In this case also, the plateau at $M=M_s/2$ in the pure chain shifts due to randomness. However, this shift is proportional to $p$ as,
\begin{equation}
\Mpc=\frac{\Ms}{2}(1+p),
\label{linear}
\end{equation}
as predicted by Cabra et al.\cite{cabra} and shown in  Fig. \ref{shiftq} by the dotted curve.

As discussed by Totsuka\cite{totsuka1}, the plateau due to STSB at $M=M_s/2$ in the model (\ref{ham1}) without randomness is formed by the many body effect. This plateau state can be regarded as a kind of Mott insulator of triplet pairs. This feature is also reflected in the nonlinear dependence of the plateau magnetization on $p$. On the other hand, the plateau state at $M=M_s/2$ of model (\ref{qmer}) with $p=0$ can be understood by the local picture. Therefore the plateau magnetization  just shifts linearly with the number of the impurity bonds.
\section{Summary and Discussion}
The magnetization plateaux in random quantum spin chains are investigated by the DMRG method. In the random mixture of two frustrated and dimerized chains  with different values of dimerization parameter, it is verified  that the plateau accompanied by STSB is destroyed by randomness even in the interval where the uniform chains have a plateau in common. It is also verified that the plateau due to imposed spatial periodicity is  robust. These are consistent with the analytical prediction by Totsuka\cite{totsukar}. The small randomness induced plateau is also predicted near the saturation field. This plateau is due to the localization of inverted spins. 

For the frustrated chains  with bond alternation and random sign strong bonds, it is found that the nontrivial plateau due to STSB splits into two smaller plateaux. The physical picture of these plateau states are clarified. In contrast to the plateaux without STSB, the magnetization on the plateaux varies nonlinearly with $p$.

So far, the experimental study of the interplay between the randomness, frustration and magnetization plateau phenomenon has not yet been carried out. Considering the variety of phenomena predicted analytically and numerically and recent progress of high magnetic field technique, fruitlful physics is expected in this field in the near future.

The computation in this work has been done using the facilities of the Supercomputer Center, Institute for Solid State Physics, University of Tokyo and the Information Processing Center, Saitama University.  This work is supported by a Grant-in-Aid for Scientific Research from the Ministry of Education, Culture, Sports, Science and Technology, Japan.  

\end{document}